\documentclass[aps,twocolumn,floatfix, showpacs,eqsecnum,      
superscriptaddress,prb]{revtex4}
\usepackage{color}
\usepackage{graphicx}
\usepackage{amsmath}
\usepackage{amssymb}
\usepackage{bm}

\begin{document}

\title{On the low-temperature anomalies in the thermal conductivity of
plastically deformed crystals due to phonon-kink scattering}
\author{J. A. M. van Ostaay}
\affiliation{Instituut-Lorentz, Universiteit Leiden, P.O. Box 9506, 2300 RA
Leiden, The Netherlands}
\author{S. I. Mukhin}
\affiliation{Theoretical physics and quantum technologies department, NITU MISIS, 119991 Moscow, Russia}
\author{L. P. Mezhov-Deglin}
\affiliation{Institute of Solid State Physics RAS, 2 Institutskaia, 142432
Chernogolovka, Russia}
\date{September, 2012}
\pacs{72.10.-d, 72.15.Eb, 66.70.-f, 61.72.Lk,67.80.-s
E-mail: vanostaay@lorentz.leidenuniv.nl}

\begin{abstract}
Previous experimental studies of the thermal conductivity of plastically
deformed lead crystals in the superconducting state have shown strong anomalies
in the thermal conductivity. Similar effects were also found for the thermal
conductivity of bent ${}^4\text{He}$ samples. Until now, a theoretical
explanation for these results was missing. In this paper we will introduce the
process of phonon-kink scattering and show that it qualitatively explains the anomalies
that experiments had found.
\end{abstract}

\maketitle

\section{Introduction}
Previous studies of the thermal conductivity of lead crystals in the
superconducting state, which were deformed plastically by low
temperature stretching of the initially perfect samples, and observation of the
recovery processes on annealing of the samples at room temperatures, had
demonstrated strong anomalies in the thermal conductivity of the
deformed Pb crystals below 4 K.~\cite{Mez79} The same effects were also seen
in weakly bent Bi crystals.~\cite{Kop73} Furthermore, experiments on the
thermal conductivity of hcp ${}^4\text{He}$ crystals grown from high pure
${}^4\text{He}$ in a long capillary had also revealed strong anomalies in
thermal conductivity of samples that were weakly deformed by bending them at
temperatures near and above 0.4 K.~\cite{Mez82, Mez84}

Several attempts for a theoretical explanation of these results have been made,
but none have unfortunately been completely succesfull.~\cite{And11} In this
primer paper however, we introduce a new model for explaning the observed
anomalies in the thermal conductivity of the weakly deformed crystals from high
pure matter. This model is based on phonon scattering on mobile
kinks on the newly induced dislocation lines. Previously, a similar model, based on scattering of
electrons by mobile kinks, has been introduced for the explanation of the anomaly 
in the electronic contribution to the thermal conductivity of plastically deformed copper 
crystals.~\cite{Muk86} In systems
where the phonon thermal conductivity is the main contribution to the transfer of heat flux, 
such as quantum crystals, metal crystals in superconducting
state and nonmetals, the scattering of thermal phonons by the mobile kinks on
dislocation lines induced under weak deformation of initially perfect samples at
reduced temperatures seems to be the natural explanation of the experimentally
observed effects. This paper will only introduce this process and show the main
results of detailed calculations of the thermal conductivity in different
directions relative to the glide plane of the dislocations. We have found that in the crystals where
scattering of phonons on kinks is the dominant scattering process our theoretical results 
can qualitatively reproduce the experimental features.
The detailed calculations referred to in this primer note and the quantitative fit of the
experimental results can be found in a paper which is soon to appear.~\cite{Ost12}

\section{Kinematics}
For a description of the kinematics of phonon-kink scattering we use a similar
procedure from Ref.~\onlinecite{Nin68}. We consider a crystal which
contains dislocations due to an external influence on the crystal. The
dislocations lie in the $xz$ plane and the direction parallel to the
dislocations is the $z$ direction. 

Around a dislocation the displacement $u_j$ can be
decomposed in two components
\begin{equation}
u_j = u^s_j + u^d_j.
\end{equation}
The "static" displacement $u^s_j$ depends on the presence of the kinks and can be
written as
\begin{equation}
u_j^s = \sum_\kappa
f_j(\mathbf{r}_{\perp}:\kappa)\xi_0(\kappa)e^{i\kappa(z-z_0(t))} +
u_{j0}^s,
\end{equation}
where $\xi_0(\kappa)$ is the Fourier transform of the dislocation's line displacement due to the kink, 
$f_j(\mathbf{r}_{\perp}:\kappa)$ is a proportionality constant and
$u_{j0}^s$ is displacement around the straight dislocation without kink. The
abbreviation $\mathbf{r}_{\perp}$ indicates $(x,y)$. The "dynamical" displacement
$u_j^d$ has its origin in the phonons and can be expressed as a superposition of
plane waves,
\begin{equation}
u_j^d = \sum_{\mathbf{k},s}q(\mathbf{k},s)e_j(\mathbf{k},s)e^{i\mathbf{k}\cdot
\mathbf{r}},
\end{equation}
where $s$ indicates the polarization of the lattice vibrations and $\mathbf{e}$
the polarization vector.
Treating the kink in a harmonic trap (potential well) with angular frequency $\Omega$ and
writing $\omega_0(\mathbf{k},s)$ for the angular frequency of the phonons
results in the total Lagrangian
\begin{subequations}
\begin{align}
L &= L_q + L_{z_0} + L_{\text{int}} + \text{const.}, \\
L_q &= \frac{\rho V}{2}\sum_{\mathbf{k},s}
\left\{\dot{q}(\mathbf{k},s)\dot{q}^*(\mathbf{k},s) \right.\nonumber \\
&\left. -\omega_0^2(\mathbf{k},s) q(\mathbf{k},s)q^*(\mathbf{k},s) \right\}, \\
L_{z_0} &= \frac{M}{2}\left \{\dot{z}^2_0(t) -  \Omega^2 (z_0(t)-z_0^0)^2
\right \} , \\
L_{\text{int}} &= -i\rho V \sum_{j, \mathbf{k}, s} k_z
\dot{z}_0(t)  e^{-ik_zz_0(t)} \xi_0(k_z) \nonumber \\
&\times F_j(\mathbf{k})\dot{q}^\ast(\mathbf{k},s) e^\ast_j(\mathbf{k},s).
\end{align}
\end{subequations}
In the equations above, $\rho$ is the density of the crystal, $V = L^3$ its
total volume, $M$ is the kink mass~\cite{Muk86}, $z_0$ indicates the position of
the kink, $z^0_0$ is its rest position and $F_j(\mathbf{k})$ is the Fourier
transform of $f_j(\mathbf{r}_{\perp}:\kappa)$, being defined as 
\begin{equation}
F_j(\mathbf{k}) \equiv \frac{1}{L^2}\int e^{-i\mathbf{k}_\perp \cdot
\mathbf{r}_{\perp}} f_j(\mathbf{r}_{\perp}:k_z) d^2r_{\perp},
\end{equation} 
where $\mathbf{k}_\perp = (k_x, k_y)$.

From the interaction term $L_{\text{int}}$ one can determine the phonon-kink
scattering amplitude per unit time $A(\mathbf{k},s; \mathbf{k}',s')$. Due to
phonon-kink scattering, phonons are no longer described by the
Bose-Einstein distribution $N^0(\omega_0(\mathbf{k},s))$. In the presence of a
small temperature gradient $\nabla T$, the linear correction to the
Bose-Einstein distribution $\delta N_{\mathbf{k}s}$ is given by
\begin{align}
& -\frac{\hbar\omega_0(\mathbf{k},s)}{k_BT^2}  ~
N^0(\omega_0(\mathbf{k},s))(1+N^0(\omega_0(\mathbf{k},s))~\nabla T \cdot
\frac{\partial \omega_0({\mathbf k},s)}{\partial {\mathbf k}} \nonumber \\
&= \sum_{s'}\int \frac{d^3 k'}{(2\pi)^3} \mathcal{P} ({\mathbf k},s;{\mathbf
k}',s') [\delta \tilde{N}_{\mathbf{k}s} - \delta \tilde{N}_{\mathbf{k}'s'}],
\label{kinetic_equation}
\end{align}
with 
\begin{equation}
\delta \tilde{N}_{\mathbf{k}s} = \frac{\delta
N_{\mathbf{k}s}}{N^0(\omega_0(\mathbf{k},s))(1+N^0(\omega_0(\mathbf{k},s)))}.
\end{equation}
and
\begin{align}
&\mathcal{P} ({\mathbf k},s;{\mathbf k}',s') = N_{ph}L^2|A
({\mathbf k},s;{\mathbf k}',s')|^2 \nonumber \\
&\times K(\omega_0({\mathbf k},s)-\omega_0({\mathbf k'},s');q_x,q_z) 
\nonumber \\
&\times N^0(\omega_0(\mathbf{k}',s'))(1+N^0(\omega_0(\mathbf{k},s))), 
\end{align}
with $N_{ph}$ the number of phonons in the crystal and 
\begin{align}
&K(\omega;q_x, q_z) = \frac{1}{L}\int dzdz'dt \exp[iq_z(z'-z) +i \omega
t] \nonumber \\
&\ll\exp[-iq_x\xi(z,0)]\exp[iq_x\xi(z',t)]\gg.
\end{align}
With Eq.~\eqref{kinetic_equation} a full kinematical treatment of the
phonon-kink scattering is possible.

\section{Heat flow}
With the full kinematics of the phonon-kink scattering at our disposal we are
able to study the effect of phono-kink scattering on the heat flow through the
crystal. The heat flux $\mathbf{Q}$ is given by
\begin{equation}
\mathbf{Q} = \sum_s\int \frac{d^3k}{(2\pi)^3} \hbar \omega_0(\mathbf{k},s)
\frac{\partial \omega_0(\mathbf{k},s)}{\partial \mathbf{k}} \delta
N_{\mathbf{k}s}  \approx
-\chi \nabla T \label{heat_flow},
\end{equation} 
where $\chi$ is the matrix of the thermal conductivity. For simplicity, we will
assume here that this matrix only has two distinct diagonal
elements and no off-diagonal elements
\begin{equation}
\chi = \begin{pmatrix}
\chi_\perp & 0 & 0 \\
0 & \chi_\perp & 0 \\
0 & 0 & \chi_\parallel        
\end{pmatrix}.
\end{equation}
This implies that there two distinct heat flows. One along the dislocation,
\begin{equation}
Q_\parallel = -\chi_\parallel (\nabla T)_z,
\end{equation}
and one perpendicular to,
\begin{equation}
\mathbf{Q}_\perp = - \chi_\perp (\nabla T)_\perp, 
\end{equation}
with $(\nabla T)_\perp = ((\nabla T)_x, (\nabla T)_y, 0)$.

Combined Eqs.~\eqref{kinetic_equation} and ~\eqref{heat_flow} allow for a full
calculation~\cite{Ost12} of $\chi_\parallel$ and $\chi_\perp$. This full
calculation shows that there are four different temperature regimes for the thermal
conductivity. These four intervals are
\begin{subequations}
\begin{align}
\text{regime 1: }& T \ll T_{\omega}, \\
\text{regime 2: }& T_{\omega} \ll T \ll T_{\Omega}, \\
\text{regime 3: }& T_{\Omega} \ll T \ll T^\ast, \\
\text{regime 4: }& T \gg T^\ast, \\
\end{align}
\end{subequations}
Here, 
\begin{subequations}
\begin{align}
T_{\omega} &= \frac{\hbar\omega_0(1/\ell)}{k_B}, \\
T_{\Omega} &= \frac{\hbar\Omega}{k_B}, \\
T^\ast &= \frac{2M\Omega^2\ell^2}{k_B},
\end{align}
\end{subequations}
where $\ell$ is the typical size of the kink and $\omega_0(1/\ell)$ is the
angular frequency for a phonon with a wavelength equal to the size of the kink.
The three temperatures are ordered as follows
\begin{equation}
T_{\omega} \ll T_\Omega \ll T^\ast.
\end{equation}

In the calculations we also took into account that in real
experiments, one does not measure the thermal conductivity in one particular
direction, but rather an average over different direction as one has no perfect
control of the orientation of the kinks. As the scattering in different
directions is a consecutive process, the scattering rates for the
different processes add. This means that the measured thermal conductivity
$\tilde{\chi}$ is found from
\begin{equation}
\tilde{\chi}^{-1} = \beta\chi^{-1}_{\perp} + (1-\beta)\chi^{-1}_{\parallel},
\end{equation}
where $\beta \in [0,1]$.

Therefore, one ends up with the following scaling behaviour for
$\tilde{\chi}^{-1}$,
\begin{equation}
\tilde{\chi}^{-1} \sim n_{ph}
\begin{cases}
\beta + n_k\mathcal{C}T^{-4} & \text{regime 1}, \\
\beta T^{-1} + n_k\left[\mathcal{C}T^{-5} +\mathcal{D}T^{-7}
\right]& \text{regime 2},\\
\beta T^{-1} + n_k\left[\mathcal{C}(1-\beta)T^{-5} +\mathcal{D}\beta T
\right]& \text{regime 3},\\
\beta T^{-1} + n_k\left[\beta T^{-3} + (1-\beta)T^{-5} \right]& \text{regime 4},
\end{cases} \label{scaling}
\end{equation}
where $n_{ph} = N_{ph}/L^2$ and $n_k = N_k/L$ are the phonon and kink
densities respectively. The script letters indicate other quantities than
the ones expressed already in the equations above.

\section{Comparison with experimental data and conclusion}
We compare our qualitative theoretical results with experimental data in
Ref.~\onlinecite{Mez79}. In figure 1 of this reference one sees that for a
sample of highly purified lead which has been plastically stretched at low
temperatures, the thermal conductivity at low temperatures has a peculiar
shape: up to certain temperature it increases with temperature, then starts
decreasing and for even higher temperatures it starts increasing with temperature
again. Annealing can make this effect less pronounced, but it seems not to be
able to completely remove this feature. Assuming that $\beta$ is neither $0$ or
$1$ and taking numerical results into account~\cite{Ost12}, one sees from
Eq.~\eqref{scaling} that for low temperatures $\tilde{\chi}$ scales as
\begin{equation}
\frac{T^4}{n_k\mathcal{C}+T^4},
\end{equation}
for higher temperatures as
\begin{equation}
\frac{T^5}{n_k\mathcal{C} + T^4},
\end{equation}
for even higher temperatures as
\begin{equation}
\frac{T^{-1}}{n_k\mathcal{C}+T^{-2}},
\end{equation}
and at the highest temperatures as
\begin{equation}
\frac{T^5}{n_k\mathcal{C} + T^4}.
\end{equation}
So at the highest and semilowest temperatures, the scaling behaviour is the
same. The exact prefactors are different of course.
This mimics the behaviour shown in the experimental data. In the semi-highest
temperature regime the thermal conductivity will decrease with
temperature, while in the other regimes the thermal conductivity
will increase with temperature. 

When comparing curves 6 and 7 in figure one, one sees that curve 6 and 7
have similar behaviour for higher temperatures. For lower temperatures though,
curve 6 lies under curve 7. As curve 6 shows the thermal conductivity
for a sample which has been deformed, while curve 7 shows the thermal
conductivity for a lead sample which has not been deformed at all, this is in
full agreement with the theory. The power-law for the thermal conductivity for a
sample with none or very little kinks has a lower power than that for a sample
with many kinks. Therefore it makes sense that for low temperature, the
thermal conductivity for a sample with many kinks is lower than that for a
sample with very little kinks. For this observation, we can therefore conclude
that samples which have not been plastically deformed at all show a much weaker
version of this effect, proving that this effect is indeed caused by phonon-kink
scattering.  This also shows that only a small amount of kinks are needed to let
this effect appear. 

The experimental data for the normal state does not match with our theoretical
calculations at all, since in the normal state the phonon contribution to the
heat flux transport is much weaker than the electron contribution. Therefore
the effect of phonon-kink scattering is not visible in that case.

We thus see that the results of our model qualitatively agree with the
experimental data. For a quantitative comparison we refer to
Ref.~\onlinecite{Ost12}.
The work of S.I. Mukhin is in part supported by RFFI grant 12-02-01018. The work
of J.A.M. van Ostaay was supported by an ERC Advanced Investigator Grant.

\end{document}